\documentclass[aps,preprint,showpacs]{revtex4}
\usepackage[dvips]{graphicx}
\usepackage{latexsym}
\usepackage{amsmath}

\newcommand{\BE}{\begin{equation}}
\newcommand{\EE}{\end{equation}}
\newcommand{\BA}{\begin{eqnarray}}
\newcommand{\EA}{\end{eqnarray}}
\newcommand{\Tr}{\mathrm Tr}

\begin{document}

\title{Grand unification in the minimal left-right symmetric 
extension of the standard model}

\author{Fabio Siringo}
\affiliation{Dipartimento di Fisica e Astronomia, 
Universit\`a di Catania,\\
INFN Sezione di Catania and CNISM Sezione di Catania,\\
Via S.Sofia 64, I-95123 Catania, Italy}

\date{\today}
\begin{abstract}
The simplest minimal left-right symmetric extension of the standard model is studied in 
the high energy limit, and some consequences of the grand unification hypothesis are
explored assuming that the parity breaking scale is the only relevant energy between the  
electro-weak scale and the unification point. While the model is shown to be
compatible with the observed neutrino phenomenology, the parity breaking scale and the heavy boson
masses are predicted to be above $10^7$ TeV, quite far from the reach of nowadays experiments.
Below that scale only an almost sterile right handed neutrino is allowed with a mass 
$M(\nu_R)\approx 100$ TeV.

\end{abstract}

 

\maketitle

An interesting aspect in the phenomenology of neutrinos is the emerging of 
important elements of new physics beyond the standard model (SM), since
there is no doubt that the experimental results
can only be understood if the neutrinos are assumed to have nonvanishing 
masses and mixings.
Massive neutrinos require the existence of a right-handed neutrino, which makes the B-L 
generator triangle anomaly free, and the related symmetry gaugeable.
Thus the most natural extension of the SM gauge group 
is the Left-Right (LR) symmetric group 
$SU(2)_L\otimes SU(2)_R\otimes U(1)_{B-L}$ which breaks to
the SM group at some high scale\cite{pati,mohapatra,senjanovic2,senjanovic}.
LR models have been discussed as embedded in larger Grand Unification (GU) models 
like SO(10), and their symmetry 
breaking path has been discussed  by several authors\cite{babu,caldwell,mohapatra2,esteves}.
Recently \cite{esteves,GU,Lindner,chavez}
there has been a renewed interest in the minimal 
L-R symmetric extension of the SM\cite{siringo,sarkar}, a model that, 
with two scalar doublets and no
bidoublets, predicts the low energy phenomenology of the SM with a very modest
cost in terms of new particles that are required to be detected at very high energy.

Despite its simple particle content,
the minimal model retains most of the interesting properties of more complex LR models:
B-L is a gauge symmetry, with a triangle anomaly free generator;
parity is spontaneously broken;
massive neutrinos can be accomodated by seesaw mechanisms;
dark matter could in part be accounted for by right handed neutrinos.

Even if at tree-level the minimal LR model, without bidoublets, fails to predict a stable broken symmetry 
vacuum\cite{siringoPRL}, a consistent path for the breaking of symmetry has been predicted
by inclusion of higher order corrections that become relevant in the Higgs-Top sector\cite{siringoPRD}.
Other paths towards the breaking of symmetry have been recently discussed\cite{GU,chavez}, 
and the minimal
model seems to be a viable first step towards new physics beyond the SM.

However, while there is a certain amount of results on the "standard" LR 
model\cite{standardLR,parida1,parida2,pal}, 
the more recent minimal
LR model has not been studied enough.
Like other non-susy models, there is evidence that an intermediate high scale is required before
unification\cite{senjanovic3}.
Thus it would be interetsing to investigate the issue of high energy unification in the
framework of the minimal LR model.

In this paper we present a detailed quantitative analysis of 
the most simple symmetry breaking path for the minimal
LR model up to unification,
in order to pinpoint its predictions for the breaking scales
and neutrino masses. While similar analysis have been presented for
more complex susy LR models\cite{patra1,patra2}, we notice
that the exact behaviour of gauge couplings depends on the detailed particle
content of the model, and it is interesting to address the question in a truly minimal
LR model with a minimum particle content.

We show that the simple hypothesis that a single intermediate scale exists
between GU and the weak scale, is enough for predicting this intermediate
scale and the masses of the heavy gauge bosons $Z^\prime$ and
$W_R$.
The prediction of a breaking scale of order $10^{10}$ GeV, halfway between the electro-weak
scale and the GU scale, is encouraging even if that scale seems to be too large
to be detected by nowadays experiments. The results are compatible with a micro-milli-eV mass 
scenario for neutrinos, and show that
the non-susy minimal LR model is a valid and natural option as a first step 
towards the understanding of new physics beyond the SM.

An interesting point is that the present analysis does not make any use of the details of
the model above GU, and does not require full knowledge of the symmetry breaking mechanism
nor detailed descriptions of the minimal set of Higgs representations: the beta functions only
depend on the actual particle content of the model below unification. This generality
makes the analysis valid for a quite large range of mechanisms and even for different unifying
groups. On the other hand, this choice of generality can be regarded as a shortcoming of the
present study, just because no answer can be given to important issues like the details on
the emerging of the low energy Lagrangian, the flow of the merged couplings above GU, the proton
lifetime prediction, and even the details of the unifying group.
Nevertheless the analysis is very simple and its generality makes it worth to be dicussed 
together with its possible effects on the physics of neutrinos.

The minimal LR symmetric model has been described in several 
papers\cite{siringo,sarkar,siringoPRL,siringoPRD}. 
The LR symmetric lagrangian is the sum of a fermionic term ${\cal L}_f$, 
a standard Yang-Mills term  ${\cal L}_{YM}$ for the gauge bosons, a Higgs term 
${\cal L}_H$ and eventually the Higgs-fermion interaction term
${\cal L}_{int}$.
A special feature of the minimal model is its limited particle content. The Higgs sector contains
two scalar doublets but no bidoublet, and is described by the simple Lagrangian  
\BE
{\cal L}_H=-{1\over 2} \vert D_L^\mu \chi_L\vert^2
-{1\over 2} \vert D_R^\mu \chi_R\vert^2+V(\chi_L,\chi_R)
\label{Lhiggs}
\EE
where the covariant derivative $D_a^\mu$ is defined according to
\BE
D_a^\mu=\left(\partial^\mu-i g_a \vec A^\mu_a \vec T_a
+i\tilde g B^\mu {Y\over 2}\right), \qquad\qquad a=L,R.
\label{covariant}
\EE
$\vec T_L$, $\vec T_R$ and $Y$ are the generators of $SU(2)_L$, $SU(2)_R$
and $U(1)_{B-L}$ respectively, with couplings $g_L=g_R=g$ and $\tilde g$.
The electric charge is given by $Q={T_L}_3+{T_R}_3+Y/2$.
The Higgs fields $\chi_a$ are the scalar doublets 
\BE
\chi_L={\chi^+_L \choose\chi^0_L},\qquad \chi_R={\chi^+_R \choose\chi^0_R}
\label{higgs}
\EE
with the trasformation properties
\BE
\chi_L\equiv(2,1,1),\qquad\qquad \chi_R\equiv(1,2,1).
\EE

A standard ${\cal L}_{YM}$ is considered for the seven gauge fields $\vec A^\mu_L$,
$\vec A^\mu_R$ and $B^\mu$.

Fermions are described by doublets of spinors $\psi_L$, $\psi_R$ with the
transformation properties 
\BE
\psi_L\equiv(2,1,B-L),\qquad\qquad \psi_R\equiv(1,2,B-L).
\EE
Their lagrangian term ${\cal L}_f$ is
\BE
{\cal L}_f=-\bar\psi_L \gamma_\mu D_L^\mu\psi_L
-\bar\psi_R \gamma_\mu D_R^\mu\psi_R
\label{Lf}
\EE

The lagrangian ${\cal L}={\cal L}_f+{\cal L}_{YM}+{\cal L}_H$ 
is fully symmetric for L-R exchange if the Higgs potential 
$V(\chi_L,\chi_R)$ is assumed to be symmetric for the exchange of
$\chi_L$ and $\chi_R$.

The simplest path for symmetry breaking requires two energy scales\cite{siringoPRD}:
parity is assumed to be broken at a large energy scale $\mu=\Lambda_R$ where
the scalar R-doublet $\chi_R$ takes a broken symmetry vacuum expectation
value (vev), $\langle\chi_R\rangle=w$ , while the L-doublet $\chi_L$ still 
retains a vanishing vev. Below this energy scale the gauge group is broken 
to the SM gauge group $SU(2)_L\otimes U(1)$. At the electroweak scale the 
L-doublet $\chi_L$ takes a broken symmetry vev $\langle\chi_L\rangle=v$,
breaking the SM gauge group to the simple $U(1)_{em}$ group of electromagnetism.
Provided that $w>>v=246$ GeV the model predicts the same phenomenology of the SM.
In unitarity gauge we set $\chi^+_a=0$ and take $\chi^0_a$ real with a finite
vev $\langle\chi^0_L\rangle=v$, $\langle\chi^0_R\rangle=w$.
Assuming $v<<w$, 
the mass matrix for the gauge bosons has two charged eigenvectors\cite{siringo}
$W^\pm_L$ and $W^\pm_R$ which are decoupled with masses
\BE
M_{W(L)}={{gv}\over 2}, \qquad
M_{W(R)}={{gw}\over 2},
\label{MW}
\EE
a vanishing eigenvalue for the electromagnetic unbroken $U(1)_{em}$ eigenvector,
and two massive neutral eigenvectors with a small mass
\BE
M_Z^2={{g^2v^2(g^2+2\tilde g^2)}\over{4(g^2+\tilde g^2)}}+{\cal O}(v^2/w^2)
\label{MZ}
\EE
for the ``light'' $Z$ boson,
and a large mass
\BE
M_{Z^\prime}^2=\left(M_{W(L)}^2+M_{W(R)}^2\right)(1+\tilde g^2/g^2)-M_Z^2
\label{MZprime}
\EE
for the ``heavy'' boson $Z^\prime$.
Of course, at low energy, all the effects of the heavy $Z^\prime$ and
$W^{\pm}_R$ are suppressed\cite{siringo}. 

In an intermediate energy range, above the electroweak scale up to the parity breaking
scale $\Lambda_R$, the minimal LR model mimics the SM with a $SU(2)_L$ gauge coupling
$g_2=g$ and a $U(1)$ coupling $g_1$ that according to Eq.(\ref{MZ}) must satisfy the
matching condition
\BE
\frac{1}{g_1^2}=\frac{1}{\tilde g^2}+\frac{1}{g_2^2}
\label{match}
\EE
at the scale $\mu=\Lambda_R$, in order to recover the known SM result $2M_Z^2/v^2=g_2^2+g_1^2$.

The GU hypothesis of a single unified gauge symmetry describing all forces and
matter at very short distances is very attractive and, according to it,
the couplings are expected to merge at a very high energy
scale $\mu=\Lambda_{GUT}$. 
However, as in other non-susy models, an intermediate high scale is required before
unification\cite{senjanovic3}. In this paper we explore the simplest hypothesis that
the intermediate scale is the parity breaking scale $\mu=\Lambda_R$, and that above that scale the
gauge couplings of the full gauge group $SU(3)\otimes SU(2)_L\otimes SU(2)_R\otimes U(1)_{B-L}$ 
run up to the unification scale $\mu=\Lambda_{GUT}$ where they merge.
We show that this simple hypothesis is enough for determining the scales $\Lambda_R$ and $\Lambda_{GUT}$,
by a simple use of the known beta functions of the model.

We prefer the use of simple one-loop beta functions that are decoupled and allow for a full
analytical discussion of the problem. Two-loop beta functions are known for the standard model\cite{arason}
and would be required for a full quantitative discussion, but their use would not change the qualitative
result in any way.

At one-loop the gauge couplings satisfy the renormalization group  (RG) equation
\BE
\mu\frac{{\rm d}g_i}{{\rm d} \mu}=\beta_i (g_i)= -b_i \frac {g_i^3}{16\pi^2}
\label{RG}
\EE
where the coefficient $b_i$ is known to be\cite{chiu4}
\BE
b_i=\frac{11}{3} C_A-\frac{4}{3} (2 n_g T_F)-\frac{1}{3} T_s n_s,
\label{b0}
\EE
$n_g$ is the number of fermion generations and $n_s$ is the number of
complex scalars.

For the gauge groups $SU(2)$ and $SU(3)$ the running of the couplings is
not affected by the breaking of symmetry at the scale $\mu=\Lambda_R$, and the
coefficients are $b_2=19/6$ for $SU(2)$ ($C_A=2$, $T_F=1/2$, $n_g=3$, $T_s=1/2$ and $n_s=1$)
and $b_3=7$ for $SU(3)$ ($C_A=3$, $T_F=1/2$, $n_g=3$ and $n_s=0$).
In fact for both  groups the trace of the square of a generator $T$ reads
\BE
\Tr{(g_i T\cdot g_i T)}= 4 g_i^2 n_g T_F=2 g_i^2 n_g
\label{Trsu}
\EE

For the $U(1)_{B-L}$ gauge group the running of the coupling depends on the particle content that
is different below and above the breaking symmetry scale.
For $\mu>\Lambda_R$ the LR symmetry is unbroken and for each generation there are six 
left-handed quarks with $Y=1/6$, six right-handed quarks with $Y=1/6$, two left handed leptons
with $Y=1/2$ and two right-handed leptons with $Y=1/2$. Thus
\BE
\Tr{(\tilde gT\cdot\tilde g T)}= \sum_{fermions} (\tilde g^2 Y^2)=\frac{4}{3}\tilde g^2 n_g
\label{Trtilde}
\EE
which is equivalent to set $T_F=1/3$ in Eq.(\ref{b0}), and since there are two scalars, $n_s=2$,
we obtain the coefficient $\tilde b=-3$ for the coupling $\tilde g$ of $U(1)_{B-L}$.
For $\mu<\Lambda_R$ the $LR$ symmetry is broken and the coupling $g_1$ is determined by
the beta function of the SM $U(1)$ gauge group: 
for each generation there are six 
left-handed quarks with $Y=1/6$, three right-handed quarks with $Y=2/3$, three right-handed quarks 
with $Y=-1/3$, two left handed leptons
with $Y=-1/2$ and one right-handed lepton with $Y=-1$. Thus
\BE
\Tr{(g_1T\cdot g_1T)}= \sum_{fermions} (g_1^2Y^2)=\frac{10}{3} g_1^2 n_g
\label{Tr1}
\EE
which is equivalent to set $T_F=5/6$ in Eq.(\ref{b0}), and since there is one scalar, $n_s=1$ (the heavy
fields are integrated out),
we obtain the SM coefficient $b_1=-41/6$ for the coupling $g_1$.

It is useful to rescale the couplings in order to make explicit the equivalence of the trace in
Eqs.(\ref{Trsu}),(\ref{Trtilde}) and (\ref{Tr1}). 
Let us define the new set of couplings
\BE
\alpha_1=\frac{5}{3} \frac{g_1^2}{4\pi}; \qquad \tilde \alpha=\frac{2}{3} \frac{\tilde g^2}{4\pi}
\label{a1t}
\EE
\BE
\alpha_2=\frac{g_2^2}{4\pi}; \qquad \alpha_3=\frac{ g_3^2}{4\pi}
\label{a23}
\EE

In fact the GU hypothesis requires that the
trace in Eq.(\ref{Trsu}) and in Eq.(\ref{Tr1}) must be the same at the GU scale $\mu=\Lambda_{GUT}$
where $SU(3)$, $SU(2)_L$, $SU(2)_R$ and $U(1)_{B-L}$ are restored as sub-groups of the same larger group.
In terms of the new set of rescaled couplings the equivalence of the trace is satisfied whenever
the couplings are equal, and the condition for GU is simply stated as
$\tilde\alpha=\alpha_2=\alpha_3$.

The new set of couplings satisfy the RG equation
\BE
\mu\frac{{\rm d} \alpha_i}{{\rm d} \mu}=\beta_i (\alpha_i)= -2 c_i \frac {\alpha_i^2}{4\pi}
\label{RG2}
\EE
where $c_2=b_2$, $c_3=b_3$, $c_1=3 b_1/5$ and $\tilde c= 3\tilde b/2$.

Eq.(\ref{RG2}) can be easily integrated yielding the linear equations
\BE
\alpha_i^{-1}(\mu)=\alpha_i^{-1}(\mu_0)+ \frac{c_i}{2\pi}\ln\left(\frac{\mu}{\mu_0}\right)
\label{flow}
\EE
that are reported in Fig.1. In this scenario the scale $\Lambda_{GUT}$ is determined by the
crossing of $\alpha_2$ and $\alpha_3$. Inclusion of two-loop corrections would decrease 
the value of $\ln \Lambda_{GUT}$ by less than $3\%$\cite{arason}, and would not affect the 
order of magnitude of $\Lambda_{GUT}$. Two-loop corrections are even smaller at the intermediate
scale $\Lambda_R$ and they are completely negligible at the electro-weak scale.

As discussed above, we assume that at the scale $\mu=\Lambda_{GUT}$ all the couplings cross, yielding
$\tilde\alpha(\Lambda_{GUT})=\alpha_2(\Lambda_{GUT})=\alpha_3(\Lambda_{GUT})$.
By the RG Eq.(\ref{flow}) the coupling $\tilde \alpha$ is let flow down to
the parity breaking scale $\Lambda_R$ where, according to the matching condition Eq.(\ref{match}),
must satisfy the constraint
\BE
\alpha_1(\Lambda_R)=\frac{5 \alpha_2(\Lambda_R) \tilde\alpha(\Lambda_R)}{2\alpha_2(\Lambda_R)
+3\tilde \alpha(\Lambda_R)}
\label{match2}
\EE
with $\alpha_1(\Lambda_R)$ that can be determined by the SM beta function Eq.(\ref{flow}) for
$\mu<\Lambda_R$, starting from the
known value at the elctro-weak scale $\alpha_1(M_Z)$ and flowing up to the
matching point $\Lambda_R$.
The unknown scale $\Lambda_R$ is pinpointed by the matching Eq.(\ref{match2}) as shown in Fig.1.

\begin{figure}[ht]
\includegraphics[height=12cm, width=9cm, angle=-90]{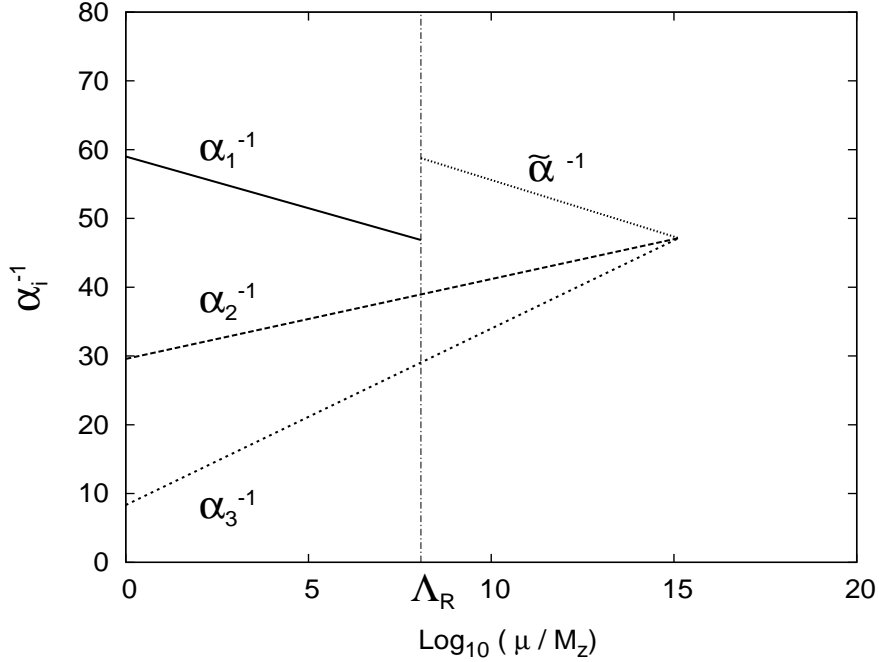}
\caption{\label{Fig1}
Running of the inverse couplings $\alpha_1^{-1}$, $\alpha_2^{-1}$, $\alpha_3^{-1}$ for
the SM $SU(3)\otimes SU(2)_L\otimes U(1)$ gauge group, 
below the parity breaking scale ($\mu<\Lambda_R$, on the left), and of the inverse couplings
$\tilde \alpha^{-1}$, $\alpha_2^{-1}$, $\alpha_3^{-1}$ for the 
LR symmetric $SU(3)\otimes SU(2)_L\otimes SU(2)_R \otimes U(1)_{B-L}$ gauge group, 
above the parity breaking scale ($\mu>\Lambda_R$, on the right).}
\end{figure}

The analytical solution is
\BE
\ln \left(\frac{\Lambda_{GUT}}{M_Z}\right)=
2\pi\left(\frac{\alpha_2^{-1}-\alpha_3^{-1}}{c_3-c_2}\right)=
\frac{12\pi}{23}\left(\alpha_2^{-1}-\alpha_3^{-1}\right)
\label{lamgut}
\EE

\BE
\ln \left(\frac{\Lambda_R}{M_Z}\right)=
2\pi\left(A\alpha_1^{-1}+B\alpha_2^{-1}+C\alpha_3^{-1}\right)
\label{lamr}
\EE

where $\alpha_i=\alpha_i(M_Z)$ are the couplings evaluated at the electro-weak scale
$\mu=M_Z$, and 

\BE
A=\left(\frac{3}{5} c_2+\frac{2}{5} \tilde c -c_1\right)^{-1}=\frac{5}{21}
\label{A1}
\EE

\BE
B=\frac{2A}{5}\left(\frac{\tilde c-c_3}{c_3-c_2}-\frac{3}{2}\right)=
-\frac{3}{7}
\label{B1}
\EE

\BE
C=\frac{2A}{5}\left(\frac{c_2-\tilde c}{c_3-c_2}\right)=\frac{4}{21}.
\label{C1}
\EE

Inserting the actual phenomenological values\cite{pdata} $\alpha_1^{-1}=59.01$,
$\alpha_2^{-1}=29.57$, $\alpha_3^{-1}=8.33$ in Eq.(\ref{lamr}) 
we obtain a parity breaking scale

\BE
\frac{\Lambda_R}{M_Z}=1.2\cdot 10^8
\label{lamrnum}
\EE
that is halfway between the electroweak scale and the GU scale that
by Eq.(\ref{lamgut}) is predicted to be 
\BE
\frac{\Lambda_{GUT}}{M_Z}=1.3\cdot 10^{15}.
\label{lamgutnum}
\EE
A scale $\Lambda_R\approx 10^7$ TeV, while far from the reach of nowadays experiments,
is in agreement with the predictions of other ``standard'' LR models\cite{parida1,parida2,pal}.
It is quite reasonable to believe that the vev of the R-scalar $\chi_R$ 
is $w\approx \Lambda_R$\cite{siringoPRD}, and that $v/w\approx 10^{-8}$. 
At the LHC energy $\sqrt{s}=14$ TeV the existence 
of tiny corrections of order $\sqrt{s}/w\approx 10^{-6}$ would be hardly detected, 
and once more a confirm of the present scenario could only come from the physics of
neutrinos.

In the minimal LR model the mass generation can be understood in terms of non-renormalizable
effective operators that are generated at low energy below the symmetry breaking scale.
Mass terms can be generated by bilinear fermionic operators that must
be coupled with Higgs bidoublets or triplets respectively for Dirac or Majorana masses, in order to
preserve gauge invariance. In the minimal model a Higgs bidoublet can be written as the product
$\chi_L \chi_R^{\dagger}$   of a $SU(2)_L$ doublet times a $SU(2)_R$ doublet, yelding a factor $vw$ in
the low energy limit, and Dirac mass terms $m_D\bar \psi_L \psi_R=\gamma_D \bar \psi_L \psi_R vw$. 
A triplet can be built up from two $SU(2)_L$ doublets (or two $SU(2)_R$ doublets) yielding
a factor $v^2$ (or $w^2$) in the low energy limit, and Majorana mass terms 
$M_L\bar\psi_L^C \psi_L=\gamma_M \bar\psi_L^C \psi_L v^2$, 
$M_R\bar\psi_R^C \psi_R=\gamma_M \bar\psi_R^C \psi_R w^2$. Here the 
couplings $\gamma_D$ and $\gamma_M$ are expected to scale like the inverse of some large
energy scale $\Lambda$.

Thus for neutrinos the mass matrix can be written as
\BE
\left(\begin{array}{cc}
M_L &   m_D  \\
m_D &   M_R  \\
\end{array}\right)= m_D
\left(\begin{array}{cc}
y\displaystyle{\frac{v}{w} } &   1  \\
1  &   y\displaystyle{\frac{w}{v}} \\
\end{array}\right)
\label{Mmatrix}
\EE 
where $y=\gamma_M/\gamma_D$ is of order unity, and the Dirac mass $m_D$ is expected to
fall in the MeV-GeV range like for other fermions. In fact for charged fermions $y=0$ and
the mass matrix contains only Dirac terms.
In the present argument a single generation is considered for neutrinos. Of course the
discussion of important 
aspects like mixing among light neutrinos would require a full mass matrix, but the existence
of mixing terms would not change in any important way the qualitative nature of the argument.  
The eigenvalues of the mass matrix Eq.(\ref{Mmatrix}) 
show the usual seesaw behaviour with a light neutrino $\nu_L$
\BE
M(\nu_L)=\frac{y^2-1}{2}\left(\frac{v}{w}\right) m_D
+{\cal O} \left(\frac{v^2}{w^2}\right)
\label{ML}
\EE
and a heavy neutrino $\nu_R$
\BE
M(\nu_R)=y\left(\frac{w}{v}\right) m_D 
+{\cal O} \left(\frac{v}{w}\right).
\label{MR}
\EE
Assuming that $v/w\approx M_Z/\Lambda_R\approx 10^{-8}$, the mass of the light neutrino 
$\nu_L$ would be pushed below the $eV$ scale while a heavy neutrino $\nu_R$ would be
conceivable with a mass $M(\nu_R)=m_D\cdot 10^8\approx 10^8$ MeV = 100 TeV.
At the LHC energy the ratio $\sqrt{s}/ M(\nu_R)\approx 0.1$, but the heavy neutrino
only interacts through the heavy gauge bosons $W_R$, $Z^\prime$ with an effective weak coupling 
that scales like $M_{W_L}^2/M_{W_R}^2=v^2/w^2\approx 10^{-16}$ compared to the light neutrino.
Thus its sterile nature would prevent its detection anyway. 
Astrophysical effects could be considered,
as the large mass of the heavy neutrino would imply important gravitational effects, 
and the dark matter of the universe could in part be accounted for by sterile neutrinos.

In summary, the simplest minimal LR extension of the SM has been studied in the high energy
limit, and some consequences of the GU hypothesis have been explored assuming that the parity
breaking scale $\Lambda_R$ is the only relevant energy above the 
electro-weak scale up to GU. In this scenario, which is shown to be
compatible with the observed neutrino phenomenology, the parity breaking scale and the heavy boson
masses are pushed up to $10^7$ TeV, quite far from the reach of nowadays experiments.
Below that scale only an almost sterile right handed neutrino could exist with a mass 
$M(\nu_R)\approx 100$ TeV.


\begin{thebibliography} {99}
\bibitem{theorem} That left-right symmetric theories predict the standard model phenomenology has been shown 
   on very general grounds by Senjanovic\cite{senjanovic} who employed the method of
   Georgi and Weinberg\cite{georgi2}.
\end{thebibliography}
\end{document}